\documentclass[reprint, 
superscriptaddress,
amsmath,amssymb,aps,
prb
]{revtex4-2}

\usepackage{graphicx}
\usepackage{dcolumn}
\usepackage{bm}
\usepackage{xcolor}
\usepackage{times}
\usepackage[normalem]{ulem}
\usepackage{units}
\usepackage[hypertexnames=false,linktocpage=true,colorlinks=true,linkcolor=blue,anchorcolor=blue,citecolor=blue,filecolor=blue,urlcolor=blue,bookmarksnumbered=true,pdfview=FitB,breaklinks=true]{hyperref}
\usepackage{float}

\usepackage{hyperref}
\begin{document}

\title{Dynamics and formation of antiferromagnetic textures in MnBi$_2$Te$_4$ single crystal}

\author{M. G. Kim}\thanks{Corresponding author}\email{mgkim@uwm.edu}
\affiliation{Department of Physics, University of Wisconsin-Milwaukee, Milwaukee, WI 53201, USA}

\author{S. Boney}
\affiliation{Department of Physics, University of Wisconsin-Milwaukee, Milwaukee, WI 53201, USA}

\author{L. Burgard}
\affiliation{Department of Physics, University of Wisconsin-Milwaukee, Milwaukee, WI 53201, USA}

\author{L. Rutowski}
\affiliation{Department of Physics, University of Wisconsin-Milwaukee, Milwaukee, WI 53201, USA}

\author{C. Mazzoli}
\affiliation{National Synchrotron Light Source II, Brookhaven National Laboratory, Upton, NY 11973, USA}

\date{\today}


\begin{abstract}
We report coherent X-ray imaging of antiferromagnetic (AFM) domains and domain walls in MnBi$_2$Te$_4$, an intrinsic AFM topological insulator. This technique enables direct visualization of domain morphology without reconstruction algorithms, allowing us to resolve antiphase domain walls as distinct dark lines arising from the A-type AFM structure. The wall width is determined to be 550(30) nm, in good agreement with earlier magnetic force microscopy results. The temperature dependence of the AFM order parameter extracted from our images closely follows previous neutron scattering data. Remarkably, however, we find a pronounced hysteresis in the evolution of domains and domain walls: upon cooling, dynamic reorganizations occur within a narrow $\sim$1 K interval below $T_N$, whereas upon warming, the domain configuration remains largely unchanged until AFM order disappears. These findings reveal a complex energy landscape in MnBi$_2$Te$_4$, governed by the interplay of exchange, anisotropy, and domain-wall energies, and underscore the critical role of AFM domain-wall dynamics in shaping its physical properties.
\end{abstract}

\maketitle

\section{introduction}

Information in real space, particularly the arrangement and evolution of antiferromagnetic (AFM) textures (domains and domain walls), may play a crucial role in determining the fundamental properties of AFM systems. For example, it has been widely recognized that topologically protected edge or surface states in intrinsic AFM topological materials can be highly sensitive to modifications of the AFM state at or near AFM domain walls (AFM DWs).\cite{1,2,3,5,6,7,8,9,10,11,12,13,14,15} Because the spin configuration at a DW differs from that inside a uniformly ordered domain, the precise AFM condition required to stabilize a topologically nontrivial state may not be fulfilled at such boundaries. Consequently, the presence, motion, or restructuring of domains and DWs can strongly affect the electronic topology. When AFM domains change in size or position, whether driven by competition with other magnetic states or external fields, topological states may be altered as well. Such modifications can manifest, for instance, as boundary-driven resistivity changes,\cite{16,17,18} providing a direct pathway to control topological properties through AFM ordering.  

\begin{figure*}[ht!]
    \centering
    \includegraphics[width=.7\linewidth]{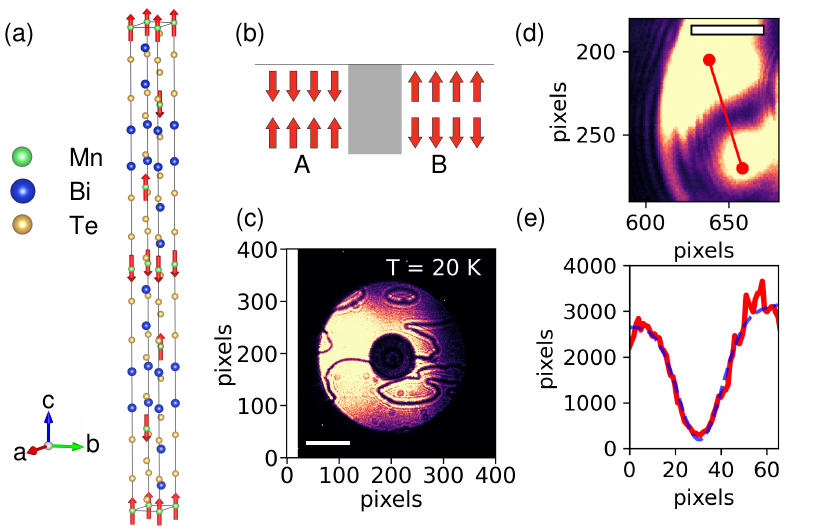}\\
    \caption{(a) Illustration of the AFM structure of MnBi$_2$Te$_4$. (b) Schematic depiction of antiphase AFM domains and the associated domain wall. The magnetic ordering phases in regions A and B differ by 180$^\circ$, representing the natural domain configuration permitted by the AFM structure in (a). The gray region between the two domains denotes the domain wall. (c) direct-CXI measurement of antiphase AFM domains and domain walls in MnBi$_2$Te$_4$, recorded at 20~K. The white bar corresponds to a scale of 20~$\mu$m. (d) Magnified view of a single domain wall. The white bar indicates 1~$\mu$m. The red line marks the trajectory of a one-dimensional cut across the wall. (e) One-dimensional intensity profile across the domain wall. The red solid line represents the experimental data, while the blue dashed line is a Gaussian fit. The extracted full width at half maximum (FWHM) of the wall is 550(30)~nm. } 
    \label{fig1}
\end{figure*}

MnBi$_2$Te$_4$ has emerged as the prototypical and first experimentally realized intrinsic AFM topological insulator (AFM TI). It is a layered van der Waals material, composed of so-called septuple layers (SLs) with the atomic sequence Te–Bi–Te–Mn–Te–Bi–Te. The magnetism originates from the Mn$^{2+}$ ions, which carry a spin $S = 5/2$ and a magnetic moment of approximately 5~$\mu_B$. Below the N\'eel temperature, $T_N \approx 24$~K, Mn moments within each SL order ferromagnetically and point out-of-plane (along $c$-axis), whereas adjacent SLs are coupled antiferromagnetically, yielding the well-known A-type AFM structure [see Fig.~\ref{fig1} (a)].\cite{MBT2,MBT3,ref57,ref55,ref56,ref58} This intrinsic configuration establishes an ideal platform for exploring the interplay between AFM ordering and topological electronic properties.  

The AFM order itself plays a central role in shaping the exotic electronic properties of MnBi$_2$Te$_4$. Angle-resolved photoemission spectroscopy (ARPES) experiments have revealed that upon cooling below $T_N$, a splitting develops in the conduction band as well as in the valence band.\cite{ref71,ref72,ref88} Furthermore, some studies have reported a Rashba-like conduction band at higher binding energies, likely of surface origin, given the expected inversion symmetry of the bulk bands.\cite{ref73} These observations highlight the importance of understanding AFM textures and their spatial variations.  

Attempts to directly probe AFM domains in MnBi$_2$Te$_4$ have been made using magnetic force microscopy (MFM).\cite{ref89,ref90} These studies have estimated the domain wall width to be approximately 500~nm and the domain size to be of order 10~$\mu$m. While MFM is highly sensitive to ferromagnetic components at the surface, such as uncompensated spins or surface spin-flop states, and has successfully revealed certain contrasts between SLs,\cite{ref90} it has inherent limitations. In particular, because MFM relies on detecting uncompensated ferromagnetic components inside DWs, its contrast diminishes when canted AFM states emerge,\cite{ref89} and it provides limited sensitivity to the subtle AFM spin configurations that dominate the bulk. Thus, despite these prior efforts, our understanding of the AFM textures in this material, namely, the detailed nature of its domains and DWs, remains far from complete.

In this paper, we investigate the AFM domains and DWs of MnBi$_2$Te$_4$ using a direct-space and real-time coherent X-ray imaging technique (direct-CXI).\cite{mgkim1,mgkim2,mgkim3} Through these measurements, we directly confirm that the antiphase domain walls, which naturally arise from the underlying A-type AFM structure, appear as distinct dark lines in the direct-space images. Furthermore, we find that the temperature dependence of the AFM order parameter obtained from our imaging results is in excellent agreement with earlier neutron scattering report. More intriguingly, however, when we carefully follow the evolution of domains and domain walls as a function of temperature, we observe highly unusual behavior: a pronounced hysteresis in domain and DW formation and dynamics. This striking reorganization of domain morphology suggests that the energy landscape of MnBi$_2$Te$_4$ is complex and that even very small changes in the balance of competing energies can significantly influence DW behavior. Our findings underscore the necessity for more comprehensive and higher-resolution investigations of AFM domains and domain walls in this system in order to fully elucidate their microscopic nature and their role in the broader physical properties of MnBi$_2$Te$_4$.

\section{experiment}

MnBi$_2$Te$_4$ single crystals were synthesized at the University of Wisconsin–Milwaukee (UWM) using the high-temperature solution growth technique, following Ref.~\cite{ref57}. High-purity Mn, Bi, and Te were combined in a molar ratio of 1:10:16, using Bi$_2$Te$_3$ as flux. The mixture was loaded into an alumina crucible, sealed under vacuum in a quartz ampoule, and heated in a muffle furnace to 900$^\circ$C at a rate of $\sim$2$^\circ$C/min. After dwelling at 900$^\circ$C for 12 hours, the melt was slowly cooled to 595$^\circ$C over two weeks and the flux was removed by centrifugation. This procedure produced large, shiny, plate-like crystals. MnBi$_2$Te$_4$ crystallizes in the trigonal $R\overline{3}m$ structure and exhibits AFM order below $T_N = 24$ K.\cite{MBT2,MBT3,ref57} 

Imaging experiments were carried out at the 23-ID-1 CSX beamline of the National Synchrotron Light Source II at Brookhaven National Laboratory. The incident X-rays were tuned to the Mn $L_3$ edge ($E = 640$ eV). A single crystal, oriented with its crystallographic $c$-axis perpendicular to the surface, was mounted on the cold finger of a liquid-helium flow cryostat integrated into a $z$-axis diffractometer configured for vertical scattering within the TARDIS UHV instrument. All measurements were performed between 15 K and 26 K. Snapshot images were collected at the AFM Bragg peak $(0,0,1.5)$ with exposure times ranging from 0.02 to 0.5 s. A Fresnel zone plate (FZP) was used to focus the X-rays, and by adjusting the sample-to-FZP distance, different magnifications were achieved.\cite{mgkim3} The variation in image size with magnification is indicated in each image by a corresponding scale bar. The scale bars were calibrated by translating the sample relative to the X-ray optics, which were mounted on a high-precision SmarAct XYZ nanoscanner assembly, thereby providing a direct measurement of the beam displacement on the sample surface.

\section{results and discussion}

MnBi$_2$Te$_4$ has been recognized as an AFM material that undergoes magnetic ordering below approximately 24 K. As illustrated in Fig.~\ref{fig1} (a), the Mn magnetic moments preferentially align along the crystallographic $c$-axis. Within each Mn layer, the spins are ferromagnetically aligned, while adjacent Mn layers couple antiferromagnetically along the $c$-axis, thereby giving rise to an overall A-type AFM configuration. The existence of such an AFM structure necessarily requires that the magnetic unit cell be doubled along the $c$-axis, leading to a magnetic propagation vector of $(0,0,0.5)$.\cite{ref57}  

In the situation where the Mn moments alternate in orientation along the $c$-axis in the characteristic $+ - + -$ sequence, the magnetic arrangement naturally gives rise to the formation of domains. Specifically, as shown in Fig.~\ref{fig1} (b), regions of opposite stacking order emerge, producing what is commonly referred to as 180$^\circ$ domains or, equivalently, antiphase domains.\cite{antiphase1,antiphase2,antiphase3} At the interface between two such regions, one encounters an antiphase DW. Across this DW, the neighboring domains differ in their magnetic phase by precisely 180$^\circ$, such that the ordering in one domain is exactly out of phase with that of the other.  

This phase difference plays a crucial role in coherent X-ray imaging measurements. In direct-CXI experiments, the destructive interference caused by this 180$^\circ$ phase shift manifests as distinct intensity minima.\cite{mgkim1,mgkim2,mgkim3} As a result, the antiphase DWs are directly visualized in the images as dark lines [Fig.~\ref{fig1} (c)]. Our observations obtained through direct-CXI are consistent with the AFM domains previously reported in magnetic force microscopy (MFM) studies.\cite{ref89,ref90}

Figure~\ref{fig1} (d) presents an enlarged image focusing on one of the antiphase DWs. By increasing the magnification through adjustment of the sample-to-FZP distance,\cite{mgkim3} a single DW was imaged in greater detail, as shown in Fig.~1(d) (scale bar = 1~$\mu$m), allowing us to examine the fine structure of the wall with higher spatial resolution. From this magnified image, one can visually estimate that the thickness of the DW is on the order of approximately 500~nm.

In order to investigate the DW in greater detail, we performed a one-dimensional analysis across the wall. The red line drawn in Fig.~\ref{fig1} (d) indicates the trajectory along which a line cut was taken, and the resulting intensity profile is displayed in Fig.~\ref{fig1} (e). In this figure, the red solid line represents the experimental data, whereas the blue dashed curve denotes a Gaussian fit to the measured profile. As is evident from the analysis, the intensity exhibits a pronounced minimum at the center of the DW, which is characteristic of destructive interference at the antiphase DW. From the Gaussian fitting, the full width at half maximum (FWHM) of the wall is determined to be 550(30)~nm. This value is in good agreement with that reported previously from MFM measurements.\cite{ref89,ref90}

\begin{figure}[!t]
    \centering
    \includegraphics[width=1\linewidth]{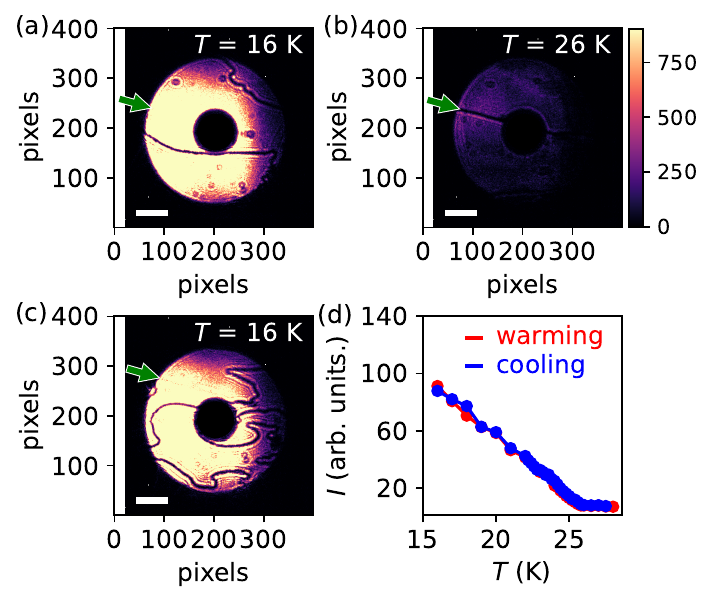}\\
    \caption{(a) AFM texture of MnBi$_2$Te$_4$ measured at 16~K after the initial cooling. Dark wavy lines correspond to antiphase AFM domain walls. (b) Direct-CXI image measured at 26~K during warming. The dark line indicated by the green arrow originates from surface-related geometrical structures and produces contrast that is independent of temperature. (c) AFM texture measured again at 16~K after recooling, showing a domain morphology distinct from that in (a). The dark lines represent antiphase AFM domain walls. (d) AFM order parameter extracted from the image intensities recorded during the thermal cycling sequence shown in (a)–(c). Red data points correspond to measurements upon warming, while blue data points correspond to those obtained during cooling. The intensity was obtained by integrating the signal over the illuminated area of images represented in (a) - (c). The AFM signal emerges below approximately 25~K. The white bar corresponds to a scale of 20~$\mu$m. The intensity scale displayed alongside panel (b) serves as the reference color scale for panels (a), (b), and (c).} 
    \label{fig2}
\end{figure}

\begin{figure*}[t!]
\centering
\includegraphics[width=.8\linewidth]{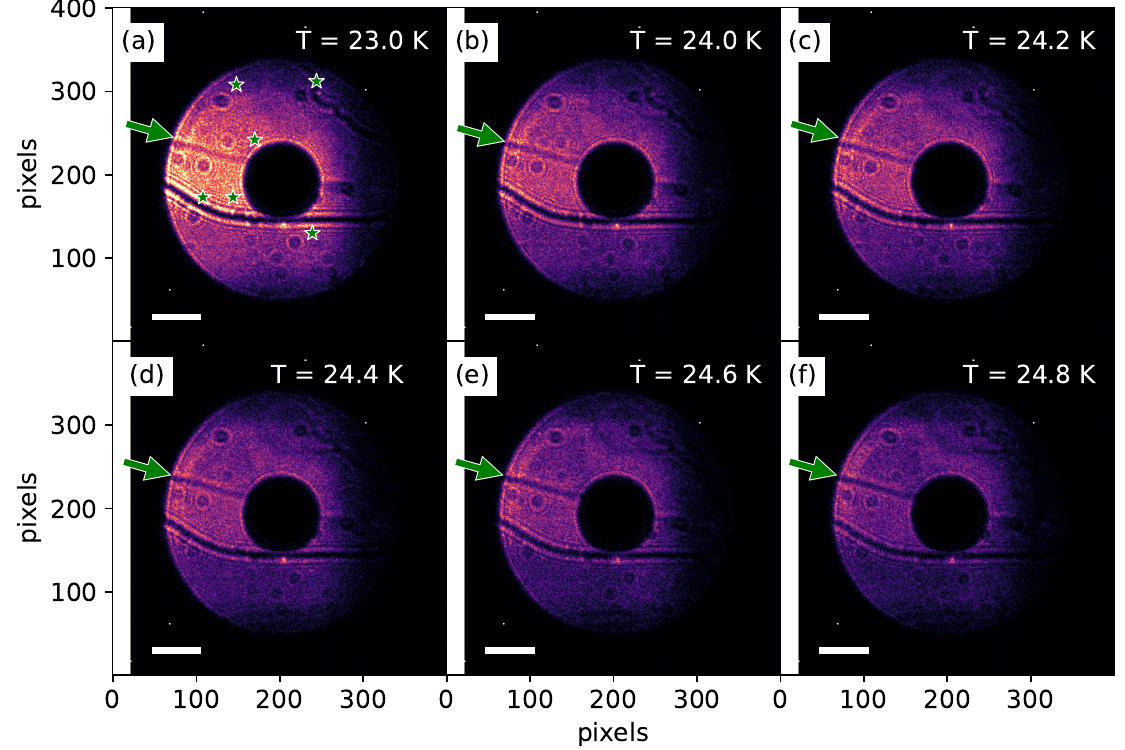}\\
\caption{Direct-CXI images of AFM textures in MnBi$_2$Te$_4$ at selected temperatures during warming. (a) Image recorded well below $T_N$ at $T = 23.0$~K. The features marked by green arrows and asterisks arise from temperature-independent geometrical structures on the sample surface and are consistently observed in all panels (b) - (f), thereby confirming that the same surface region was probed throughout. (b) $T = 24.0$~K, (c) $T = 24.2$~K, (d) $T = 24.4$~K, (e) $T = 24.6$~K, and (f) $T = 24.8$~K just below $T_N$. In these images, the curved line represents antiphase AFM domain wall. The white scale bar corresponds to 20~$\mu$m. The color map used for these images is identical to the color scheme employed in Fig.~\ref{fig2}} \label{fig3}
\end{figure*}

\begin{figure*}[t!]
\centering
\includegraphics[width=.8\linewidth]{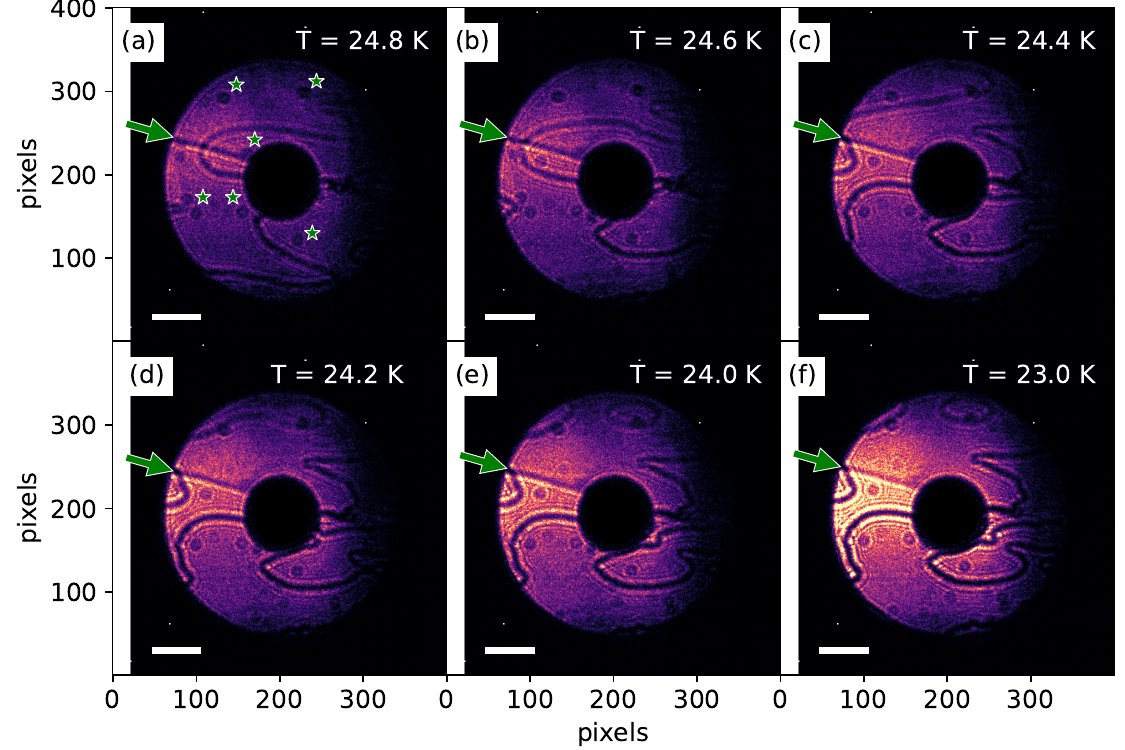}\\
\caption{Direct-CXI images of AFM textures in MnBi$_2$Te$_4$ at selected temperatures during cooling. (a) Image recorded just below $T_N$ at $T = 24.8$~K. The features marked by green arrows and asterisks arise from temperature-independent geometrical structures on the sample surface and are consistently observed in all panels (b) - (f), thereby confirming that the same surface region was probed throughout. (b) $T = 24.6$~K, (c) $T = 24.4$~K, (d) $T = 24.2$~K, (e) $T = 24$~K, and (f) $T = 23$~K. In these images, the dark wavy lines represent antiphase AFM domain walls. The white scale bar corresponds to 20~$\mu$m. The color map used for these images is identical to the color scheme employed in Fig.~\ref{fig2}} \label{fig4}
\end{figure*}

Figures~\ref{fig2},~\ref{fig3}, and ~\ref{fig4} provide a summary of the evolution of the AFM domain images as a function of temperature. Domain images were systematically acquired while varying the sample temperature. Figure~\ref{fig2} (a) displays the AFM domain pattern recorded at 16~K after the initial cooldown, serving as a reference state for subsequent temperature-dependent measurements. Upon warming the sample, similar images were collected at progressively higher temperatures and we observed no change in the domain morphology, see Fig.~\ref{fig3}. At temperatures above the N\'eel transition, the AFM DWs that were clearly visible in Fig.~\ref{fig2} (a) vanish, and only the dark lines indicated by the green arrows in Fig.~\ref{fig2} (b) remain. This dark line, in contrast to AFM DWs, originates from geometrical features on the sample surface.\cite{mgkim3,structurefeature} As such, this remains unchanged across the entire temperature range. Furthermore, at low temperatures, where the AFM signal becomes strong, the contrast arising from such surface-related geometrical structures becomes comparatively weak, rendering them nearly invisible in the images [see Fig.~\ref{fig2} (c) and Fig.~\ref{fig3} and ~\ref{fig4}]. Importantly, the temperature-independent nature of these structural features provides a robust internal reference, confirming that the same surface region of the sample was consistently probed throughout the entire series of measurements.

When the temperature is subsequently lowered, AFM domains reappear but with altered configurations, as exemplified in Fig.~\ref{fig2} (c). A direct comparison of Figs.~\ref{fig2} (a) and (c), both measured at 16~K on nearly the same region of the sample surface, clearly demonstrates that the domain patterns formed in different thermal cycles are distinct. By plotting the temperature dependence of the intensity associated with these domain images, we obtained the AFM order parameter as a function of temperature, which is shown in Fig.~\ref{fig2} (d). The intensity shown in Fig.~\ref{fig2} (d) was obtained by integrating the signal over the illuminated (donut-shaped) area of images represented in Figs.~\ref{fig2} (a) - (c). As seen in Fig.~\ref{fig2} (d), the AFM signal emerges below approximately 25~K, in agreement with previously reported results. Such temperature-dependent reconfiguration of AFM domain patterns is consistent with the general behavior of domains and DWs in the absence of strong pinning centers. In other words, when DWs are not immobilized by crystal imperfections, impurities, or geometrical defects,\cite{antiphase3,pinning1,pinning2,pinning3,pinning4} the domains tend to form and reorganize freely and randomly, in line with the conventional understanding of AFM domain formation.

Figure~\ref{fig3} and ~\ref{fig4} illustrates the remarkable process of domain and DW formation. Interestingly, during the process of domain and DW formation, we observed that the system exhibits pronounced and highly dynamic behavior when the temperature is lowered [Fig.~\ref{fig4}]. In stark contrast, however, during the warming process no such dynamic rearrangements were detected, and the domain patterns remained essentially static [Fig.~\ref{fig3}]. Figure~\ref{fig4} (a) shows the AFM domain structure imaged at $T$ = 24.8~K, a temperature slightly below the N\'eel temperature. As discussed previously, one can observe both the geometrical surface-related features (indicated by the green arrow) and the curved AFM DWs. In addition, small dot-like features, marked by asterisks, are also visible. These arise from temperature-independent surface-related structural features, such as dirt particles or impurities. At $T$ = 24.8~K, the X-ray resonant signal associated with AFM order is relatively weak [see Fig.~2(d)]; therefore, both the geometrical features and the AFM DWs are simultaneously visible in the image.  

Upon gradually lowering the temperature, we observed that the morphology of the domains and DWs underwent dramatic and abrupt changes. Comparing Figs.~\ref{fig4} (a), (b), and (c), one immediately notices that, despite a temperature difference of only 0.4~K, the domain and domain-wall patterns are strikingly different. Such rapid and drastic transformations of the domain morphology continue as the sample is cooled further, down to approximately 24~K, as shown in Figs.~\ref{fig4} (d) and (e). Upon reaching $T$ = 24~K, however, the changes appear to cease, and the domain and domain-wall configurations stabilize. This persistence is clearly demonstrated in Fig.~\ref{fig4} (f), where the domain structure measured at $T$ = 23~K retains the same pattern. Thus, within a narrow temperature interval of merely 1~K below the N\'eel temperature, the AFM domains and DWs undergo a remarkable reorganization. Notably, this dramatic reconfiguration was never observed upon warming. To restate, when the temperature was increased from $T$ = 23~K, the domain and domain-wall patterns observed in Fig.~\ref{fig4} (f) remained unchanged all the way up to the N\'eel temperature, at which point the system entered the paramagnetic state and the domains and DWs vanish.

\begin{figure}[!t]
    \centering
    \includegraphics[width=.8\linewidth]{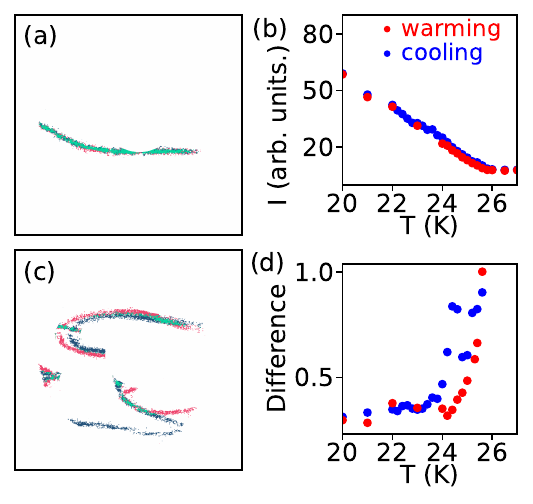}\\
    \caption{(a) and~(c). Comparison between extracted domain wall features at 24.6~K and~24.8~K during warming and cooling, respectively. Corresponding original images are shown in Fig.~\ref{fig3} and ~\ref{fig4}. (b) AFM order parameter obtained from Fig.~\ref{fig2}. (d) Difference between the images taken during warming and cooling. See text for details.} 
    \label{fig5}
\end{figure}

To examine the phenomena described above in greater detail, we compared the domain patterns imaged at different temperatures.  Figures~\ref{fig5} (a) and (c) show overlapping images obtained during warming and cooling, respectively, where the comparison is made between data taken at 24.8~K and 24.6~K as representative examples. The domain-wall features were extracted in pixels, and a pixel-by-pixel matching analysis was performed. In the composite images in Figs.~\ref{fig5} (a) and~\ref{fig5}(c), green pixels correspond to DW pixels that coincide between the two images, whereas blue (24.8~K) and red (24.6~K) pixels indicate unmatched features. From this comparison, we computed the difference ratio of unmatched pixels to the total number of DW pixels, as plotted in Fig.~\ref{fig5}(d), and compared it with the AFM order parameter shown in Fig.~\ref{fig5}(b).

As seen in Fig.~\ref{fig3}, the domain pattern remains unchanged during warming. Correspondingly, the difference ratio in Fig.~\ref{fig5} (d) is approximately~0.35 below 24 K, reflecting small numerical discrepancies that arise from the pixel-based comparison method. Although two images may appear nearly identical to the eye, slight variations in pixel intensity lead to nonzero differences in the computed ratio. Furthermore, while no change in domain-wall morphology is observed between 24~K and $T_N$ [Fig.~\ref{fig3}], Fig.~\ref{fig5} (d) shows a gradual increase in the difference ratio above~24~K. This increase is likely an artifact originating from the weakening of AFM ordering as the temperature approaches $T_N$, which reduces the signal strength of the domain walls and thereby amplifies pixel-level discrepancies. In contrast, the difference value approaching~1 at $T_N$ is considered genuine, as it represents the disappearance (or emergence) of the AFM domain pattern at the transition.  

Upon cooling, as shown in Fig.~\ref{fig5} (d), the domain patterns begin to appear just below $T_N$, where the difference value approaches~1, consistent with the sudden formation of AFM domains. With gradual decrease of temperatures, the progressive evolution of domain-wall configurations, such as those observed in Fig.~\ref{fig4}, is well captured by the variation in the difference ratio approximately between 0.6 and 0.82. Below~24~K, the difference value returns to approximately~0.35, indicating that the domain morphology becomes stable and no longer changes with temperature. When compared with the AFM order parameter shown in Fig.~\ref{fig5} (b), no corresponding anomalies are found in the same temperature range (24 K $\leq T \leq T_N$). The order parameter follows the typical behavior expected for a second-order transition, confirming that the domain-formation dynamics observed here are not driven by fluctuations in the order parameter itself.

The formation of domains and DWs observed here is rather unusual. As the temperature varies, the system explores different minima of the energy landscape. In the case of AFM ordering, the dominant competing energies expected to govern this landscape are thermal energy, exchange energy, anisotropy energy, and domain-wall energy.\cite{antiphase3,energy1,energy2,energy3,energy4} Upon cooling, thermal energy decreases, spin fluctuations are suppressed, and exchange and anisotropy energies begin to dominate the overall energy balance. This drives the onset of AFM ordering and initiates the formation of domains. Because stray fields are negligible in antiferromagnets,\cite{antiphase3,energy2} the formation and stability of domains and DWs are largely governed by domain-wall energy. The greater the number of DWs, the larger the associated energy cost; thus, as the system cools, it naturally reduces the number of DWs in order to minimize energy and achieve stability. This process leads to an overall coarsening of the domains and results in the dynamic motion, merging, or rearrangement of DWs, consistent with the behavior that we have directly observed here. 

An intriguing aspect of our results is that the dynamic motion of domain walls ceases around 24~K, suggesting that at this temperature the system attains a locally minimized configuration governed primarily by the domain-wall energy, even though other competing energy terms may not yet be fully minimized. Another possible mechanism is that the domains become frozen. In conventional spin freezing, one would expect the dynamics to follow a similar second-order–like evolution in the absence of any external perturbation (such as a magnetic-field variation). It is therefore difficult to reconcile the asymmetric domain dynamics observed upon heating and cooling with the behavior expected if a hidden mechanism analogous to spin freezing were responsible.

Upon warming, thermal energy increases and spin fluctuations become enhanced. These fluctuations are expected to influence the spins constituting the DWs as well, potentially causing the walls to broaden or become unstable. Consequently, one might anticipate thickness variations of the DWs or even the disappearance of DWs at elevated temperatures. Nevertheless, within the resolution limit of our experiments, and in agreement with earlier MFM studies\cite{ref89,ref90}, we did not detect measurable changes in domain-wall thickness nor did we observe their disappearance. This implies that the enhanced thermal energy upon warming primarily competes with the exchange and anisotropy energies, rather than exerting a direct influence on the domain-wall energy. Such competition may have influenced the underlying exchange and anisotropy energy landscape. Given the intimate connection between DWs and these energies (DW width $\delta \propto \sqrt{A/K}$ where $A$ is the exchange parameter and $K$ is the anisotropy constant~\cite{dw1,dw2,dw3,dw4,dw5}), it remains possible that the DWs are also affected; but, such modifications may be exceedingly subtle and difficult to resolve experimentally. These considerations highlight the need for more precise measurements of domain-wall evolution, as well as systematic investigations of the temperature dependence of exchange and anisotropy energies, in order to fully elucidate the hidden microscopic mechanisms at play.

\section{Conclusion}

In summary, we have investigated the formation and dynamics of magnetic domain and DWs of MnBi$_2$Te$_4$ single crystals using direct-space and real-time coherent X-ray imaging. Our measurements directly visualized AFM domains and antiphase DWs, whose widths were determined to be 550(30)~nm, consistent with earlier MFM studies. Temperature-dependent imaging revealed that the AFM order parameter follows previous reports, yet the domain morphology exhibits striking hysteresis. Specifically, while warming across the N\'eel temperature leaves the domains essentially unchanged until they vanish in the paramagnetic state, cooling induces a highly dynamic reorganization of domains within a narrow $\sim$1~K interval below $T_N$.  

This asymmetric domain evolution highlights the complex energy landscape of MnBi$_2$Te$_4$, in which thermal, exchange, anisotropy, and domain-wall energies compete to stabilize the system. Upon cooling, the reduction of thermal fluctuations favors coarsening of domains through domain-wall motion and merging, whereas upon warming, the increased thermal energy does not significantly affect the DWs, but rather competes primarily with exchange and anisotropy energies. Our findings emphasize the role of DW energetics in AFM ordering and provide new insight into the metastability and dynamics of domains in MnBi$_2$Te$_4$. Future high-resolution measurements of DW motion, together with systematic studies of the temperature dependence of exchange and anisotropy parameters, will be essential to fully establish the microscopic mechanisms underlying these phenomena.

\begin{acknowledgments}
This work was supported by the University of Wisconsin-Milwaukee. This research used resources at the 23-ID-1 beamline of the National Synchrotron Light Source II, a DOE Office of Science User Facility operated for the DOE Office of Science by Brookhaven National Laboratory under Contract No. DE-SC0012704. The resources made available through BNL/LDRD\#19-013 are acknowledged.

\end{acknowledgments}

\bibliography{MBT} 

\end{document}